\documentclass[11pt]{article}
\usepackage{latexsym}
\usepackage{amssymb}
\usepackage{amsmath}

\usepackage{hyperref}
\usepackage{pstricks}

\usepackage{amssymb,amsmath}
\usepackage{epsfig}
\usepackage[notref,notcite]{showkeys}
\usepackage{graphicx}

\newcommand{\del}{\partial}
\newcommand{\tr}[1]{\:{\rm Tr}\,#1}
\def\be{\begin{equation}}
\def\ee{\end{equation}}
\def\beqa{\begin{eqnarray}}
\def\eeqa{\end{eqnarray}}
\newcommand{\eqn}[1]{(\ref{#1})}
\newcommand{\bra}{\langle}
\newcommand{\ket}{\rangle}
\newcommand{\TT}{{\mathbb T^{\gamma}}}
\newcommand{\jj}{{\mathbb P}}
\newcommand{\rd}{\mathrm{d}}
\newcounter{lst}

\begin{document}
\begin{titlepage}

\begin{center}

\baselineskip=24pt

{\Large\bf Aspects of Group Field Theory}

\baselineskip=14pt

\vspace{1cm}

{Patrizia Vitale}
\\[6mm]
{\it Dipartimento di Scienze Fisiche, Universit\`{a} di Napoli  Federico II}\\
{\sl and INFN, Sezione di Napoli}\\
{\it Monte S.~Angelo, Via Cintia, 80126 Napoli, Italy}
\\[4mm]
 {\it Laboratoire de Physique Th\'eorique,\\
Universit\'e Paris Sud XI, Orsay, France}
\\{\small\tt
patrizia.vitale@na.infn.it}

\end{center}

\vskip 2 cm

\begin{abstract}
I review the basic ingredients of discretized gravity which motivate the introduction of
Group Field Theory. Thus I describe the GFT formulation of some models and conclude with
a few remarks on the emergence of noncommutative structures in such models.\\
\\
\textbf{keywords}: {Quantum Gravity, Noncommutative Geometry, Star Products} 
\end{abstract}

\end{titlepage}
\section*{Introduction}
This article is based on a talk given at the XX International Fall Workshop on Geometry
and Physics in  Madrid and it is aimed at illustrating the deep geometric roots of the
Group Field Theory (GFT) approach to Quantum Gravity, together with the recent emergence
of Noncommutative Geometry into the game.

The first part, partly based on \cite{Vitale2011}, is dedicated to a review of the
discrete formulation of gravity as a BF theory with constraints, including the Holst
formulation \cite{holst} with the Barbero-Immirzi parameter \cite{Barbero, Immirzi}. It
is not meant to be exhaustive while it is directed to a public of non-specialists, with
the accent put on geometric structures. I then introduce the GFT description of the
discrete BF path integral and briefly review the implementation of constraints.

In the final part I discuss the emergence of noncommutative structures at various levels
of the models considered.

\section{Gravity as a BF theory with constraints}
The starting point of this analysis is the well known reformulation of the
Einstein-Hilbert action $ S\left[\,g\,\right] =  \int_M d^4 x\ \sqrt{|g|}\ {R}
$
in the first order formalism, as
 \be
  S\left[\,e\,,\,\omega\,\right] = \int_M \tr_1\left[\ \left(e
\wedge e\right)\ \wedge F\ \right] = \int_M \frac{1}{2}\,\epsilon_{ABCD}\,e^A\wedge e^B
\wedge F^{CD} \;\; \label{action}
\ee
with $ A,B,C,D=1,..,4$. $F = D\,\omega$ is the curvature of the connection one-form of
the  principal $SO(3,1)$-bundle\footnote{I will indicate  Lie groups with capital
letters, Lie algebras with lower-case letters, a generic Lie group with $G$ and a
generic Lie algebra with $\mathcal{L}$} while {$e$} are the tetrad one-forms implicitly
defined by {$\;\; g_{\mu \nu}= e_\mu^A e_\nu^B \eta_{AB} $}. The connection one-form
$\omega$ and the tetrads  $ e$ are to be regarded as independent variables. The
equations of motion may be seen to be  equivalent to  Einstein's equations when the
tetrads are non-degenerate. Let me recall that $e= e^A_\mu dx^\mu v_A $ is  a vector
valued one form of the associated vector bundle, so that, once  a trivialization chosen
$\;  {e\wedge e = e^A_\mu e^B_\nu dx^\mu\wedge dx^\nu {\tau_{AB}}}
 \;$
  with {$\tau_{AB}\in so(3,1)$}. Moreover, the connection and the  curvature are locally of the form
 $\; \omega= \omega^{AB}_\mu dx^\mu\tau_{AB}\; $
  $\; F= F^{AB}_{\mu\nu} dx^\mu\wedge dx^\nu {\tau_{AB}}.
 $
   Therefore ${\tr_1}$ is to be intended as the  bilinear nondegenerate form on the $so(3,1)$ algebra
$$
{<J_a,P_b>_1=\eta_{ab}\; <J_a,J_b>_1=<P_a,P_b>_1=0 \;\; a,b=1,..,3}
$$
with  $J_a = J_A=\epsilon_{A}^{BC}\tau_{BC} \; \;A,B,C=1,..,3$\;\;\;\;\;\; $P_a=
P_A=\tau_{A,4}, \; A=1,..,3$.

The action \eqn{action} my be recast into the form of a BF action with constraints. The
BF action
\be
 S\left[\,B\,,\,\omega\,\right] = \int_M \tr_1\,(B\wedge F)
\ee
encodes  a topological gauge theory described in terms of   Lie algebra valued 2-forms,
$B,F$, where $F$ is the curvature. The equations of motion simply state that the
connection $\omega$ is flat and $D_\omega B=0$. The constraints ${C}(B)$ have to impose
that the $B$ field be simple, that is ${ B = e\wedge e}$. The action becomes then
\be
S_1\left[\,B\, ,\, \omega\, ,\, \phi\,\right] = \int_M\, B^{AB} \wedge F_{AB} +{C}(B).
\ee
In $so(3,1)$ we can define another trace
 $\tr_2=<.,.>_2$,
$$ <J_a,J_b>_2=<P_a,P_b>_2=\delta_{ab}, \;\; <J_a,P_b>_2=0
$$
so that  a new action is produced
\be
 S_2= \int_M \tr_2(e\wedge e\wedge
F)=\int_M e^A\wedge e^B \wedge F_{AB}[\omega].
\ee
When added to the action $S_1$ it doesn't change the equations of motion but it has
quantum consequences.
 The full action
$
  S_H= S_1+ \frac{1}{\gamma}S_2
 $
 is  the Palatini-Holst action \cite{holst};  the real parameter $\gamma$
is the Barbero-Immirzi parameter \cite{Barbero, Immirzi}. The relevance of this second
term in the action is well known: it allows to introduce the Ashtekar connection which
is at the basis of the canonical quantization programme for  gravity.

 $S_H$  may be regarded as a constrained BF action
\be
 S_H=\int_M B^{AB}\wedge F_{AB} + {C}(B) \label{sh}
\ee
 where the constraint has to implement
\be B^{AB}=\epsilon^{AB}_{CD}e^C\wedge e^D+\frac{1}{\gamma} e^A\wedge e^B.\label{constr} \ee
\subsection{Gravity in three dimensions }
In three space-time dimensions the Einstein-Hilbert action becomes in the first order
formalism
\be {S[\omega,e]=\int_M \tr\,( e\wedge F)} \;\;\;
\ee
with
\begin{itemize}
\item
$\omega= \omega^A_\mu dx^\mu {\tau_A}$, the $SO(2,1)$ connection one-form\;\; \;
{$\tau_A\in so(2,1)$}
\item {$F= F^A_{\mu\nu}dx^{\mu\nu} {\tau_A} $} the curvature of the
connection one-form
\item { $e=e^A_\mu dx^\mu {\tau_A}$} triads, with the
identification $so(2,1)\simeq V(M)$
\item $\tr \leftrightarrow$ Killing form in $so(2,1)$
\end{itemize}
This is the BF action for the gauge group $SO(2,1)$, with $B=e$. Because the $B$ field
 is a one-form there are no extra constraints to be imposed.

 We can build $BF$ models in any space-time dimensions,
 with gauge group the Lorentz group $SO(D-1,1)$ .
  For $BF$ in $D$ dimensions $F$ is always a Lie algebra valued 2-form (the
curvature), while  $B$ is a $D-2$ Lie algebra valued form. In particular
 in $D=2$  (with $B$  a zero form)\footnote{Notice however that in $D=2$ the BF formulation requires the gauge group to be the Poncar\'e or the   De Sitter group  (for details see for example \cite{Cangemi})} and $D=3$ (with $B$ a one-form), the BF action reproduces the gravity
 action and it is a topological theory. In  $D\ge 4$ BF {+ constraints} reproduces gravity, and it is
dynamical.
\subsection{Discretization of the BF action}
From now on we stick to  the  Riemannian case so that the Lorentz group is replaced by
the rotation group and we often work with its covering group. We consider space-time
 triangulations with a $D$-dimensional  simplicial complex
 $K_D=\{\sigma_D,...,\sigma_0\}$.
 To discretize  $B$ which is a Lie algebra valued $D-2$ form, we integrate it on a $D-2$ simplex
$$B\longrightarrow E\in so(D),\;\;E=\int_{\sigma_{D-2}} B.$$
To discretize the curvature 2-form we follow the prescription of   Regge calculus where,
in $D=2$ the curvature is measured by the deficit angle when turning around a vertex
(0-simplex) {$$ \delta(v)= 2\pi-\sum_{\ell,\ell'\supset v} \theta_v(\ell,\ell').$$ }
Analogously, in $D=3$ the curvature is measured
 by the deficit angle when turning around an edge (1-simplex).
Therefore, in  $D$ generic the local curvature on the triangulated manifold is detected
by the { holonomy of the connection} around a $D-2$ simplex. The closed path around the
$D-2$ simplex is the boundary of a face, $f_*$  in the { dual discretization} $K_*$.
Therefore we have
  $
  h_{\ell_*}= P\exp\int_{\ell_*\subset \del f_*} \omega
  $
  and \be
 H_{f_*} \equiv H_{\sigma_{D-2}}= \prod_{\ell_*\subset \del f_*} h_{\ell_*}
\; \ee where we have explicitly indicated the duality between   dual faces and $D-2$
simplices. The two-dimensional  sub-complex contained  in $K_*$ draws a graph
{$\mathcal{G}$}. {Each assigned $\mathcal{G} \subset |K_*|$ represents a specific
discretization of space-time}.

The discretized BF action becomes
\be
S(E_{\sigma_{D-2}}, H_{\sigma_{D-2}})=\sum_{{\sigma_{D-2}}\in K} \tr E_{\sigma_{D-2}}
H_{\sigma_{D-2}}. \ee from which we derive the  discretized partition function
\be
A[K,K_*] =  \int_{ \mathcal{L}}\prod_{{\sigma_{D-2}}\in K} d E_{\sigma_{D-2}}\int_{G}
\prod_{\ell_*\in K_*} dh_{\ell_*} \exp[{i}\tr(E_{\sigma_{D-2}} \prod_{\ell_*\in \del
f_*}h_{\ell_*})]
\ee
where $\mathcal{L}$ is the Lie algebra of the appropriate rotation group $G$.
 The
integral in the Lie algebra can be formally performed and we get
\be
A[K_*]=\int\prod_{\ell_*\in K_*} dh_{\ell_*} \prod_{f_*\in K_*}\delta(\prod_{\ell_*\in
\del f_*}h_{\ell_*}) \label{bfampl}
\ee
 This result, valid in any dimension,  is  expressed solely in terms of
the dual discretization.   It can be interpreted as the amplitude of the graph
$\mathcal{G}$ drawn in $K_*$. It is interesting to notice that the same result is
obtained independently in the  spin-foams approach \cite{spinfoams},   as the transition
amplitude from a space geometry to another.

The natural question which arises is:   what is, if any, the  field theory which
generates such Feynman graphs? Group field theory, introduced in the early 90's in
\cite{boul} and later developed by \cite{Ooguri, Freidel,Oriti} is a candidate to that.

\section{Group Field Theory}
 Group Field Theories are  a
particular family of tensor models where the fields are tensors defined on the Lorentz
group manifold. Tensor models are in turn the natural generalization of matrix models to
higher dimensions, aimed at describing  aleatory space-time geometries (for an up to
date review on the subject and recent achievements see \cite{gurau}).
 As in  more general tensor models, GFT encode  the space-time dimension in  the order of the tensor-field
 while, specific to these models,  the field arguments live on products of the Lorentz (rotation) Lie
group
$$
 \phi:  (g_1,...g_D)\in [SO(D)]^D\rightarrow \phi(g_1,...g_D).$$
   Feynman amplitudes of a $D$ dimensional GFT are  dually associated with a discrete
space-time via a specific triangulation and gluing rules given by the propagator and
vertices of the theory. The functional integral formalism defines  a weighted sum over
triangulations with each weight (amplitude) related to a sum over geometries, therefore
achieving a desired feature of any candidate quantum theory of gravity - a sum over both
topologies and geometries.

The simplest GFT models generate amplitudes of the BF type, as in \eqn{bfampl} . They
are therefore  topological models. A possible choice is to include  the dynamics by
implementing the constraints on the propagator, while the vertex of the theory would
remain unchanged with respect to the topological theory. The other possibility is to
constrain the vertex and leave the propagator unmodified. Following \cite{KMRTV2010} I
adopt here the first point of view .

The propagator $C$, is a Hermitian operator with Hermitian kernel \;$C(g_1, \ldots, g_D;
g'_1, \ldots, g'_D)$:
\be
[ C\phi ] (g_1, \ldots , g_D) = \int  d g'_1  \ldots d g'_D C(g_1, \ldots, g_D; g'_1,
\ldots, g'_D) \phi (g'_1, \ldots , g'_D).
\ee
It is represented graphically as a stranded line with $D$ strands and the precise form
of $C$ characterizes  the different models.  The vertex is the same for all models: its
kernel   is a product of  delta functions matching strand arguments, so that each delta
function joins two strands in two different lines. For instance, in three dimensions the
SU(2) BF vertex is expressed as
\beqa
 S_{{\mathrm{int}}}[\phi]&=&
\frac{\lambda}{4}\int \left(\prod_{i=1}^{12}d g_i\right)\phi(g_1,g_2,g_3)
\phi(g_4,g_5,g_6)\phi(g_7,g_8,g_9)\nonumber\\
&&\phi(g_{10},g_{11},g_{12})\;
 K(g_1, .. g_{12}),
\eeqa
with
\be
 K(g_1, .. g_{12}) = \delta(g_3g_4^{-1})\delta(g_2g_8^{-1})\delta(g_6g_7^{-1})\delta(g_9g_{10}^{-1})\delta(g_5g_{11}^{-1})\delta(g_1g_{12}^{-1})
\label{3vertex}
\ee
 It represents the gluing of four triangles to form a tetrahedron, the elementary
space-time  block in 3D. In D dimensions it is therefore replaced by a term proportional
to $\phi^{D+1}$. The propagator represents instead the gluing of two $D-$simplices along
a common face.
\subsection{ GFT for  BF theories}
The propagator for BF theories is just the projection on gauge invariant fields,
\be
\jj (\phi)=\int_{SO(D)} d h \phi(g_1 h,\ldots, g_D h),
\ee
It verifies   $\jj ^2=\jj$ so that the only eigenvalues are $0$ and $1$. This is another
manifestation of the fact that  BF models have no dynamical content. The operator  $\jj$
is Hermitian with kernel
\be
\jj (g_1, ..., g_D ; g'_1, ... g'_D)=\int d h \prod_{i=1}^D \delta (g_i h
(g'_i)^{-1})\label{prop}.
\ee
The full GFT action for these models may be  synthetically represented as
\be S[\phi]=\int dg
\phi^2+\lambda \int dg \phi^{D+1} \label{GFTaction}
\ee
where the fields are functions of D copies of the group and the integration is performed
on as many copies of the group as needed, so that $dg$ stands for the appropriate power
of the Haar measure.
 To compute the amplitude of a
given graph we assign to each propagator the definition in Eq. \eqn{prop} and to each
vertex the appropriate generalization of Eq. \eqn{3vertex}. We choose for simplicity
graphs with no external legs. After integration over all group variables associated to
the strands of propagators we obtain
\be
A_{\mathcal G} = \int \prod_{\ell_ \in {L}_{{\mathcal G}} } d h_\ell \prod_{ f \in
\mathcal{F}_{{\mathcal G}}} \delta \left( {\vec {\prod}_{\ell \in f} h_\ell^{\eta_{\ell
f} }}   \right), \label{gftampl}
\ee
where we have omitted the star labeling the dual discretization. The incidence matrix
$\eta_{\ell f}$    has value $+1$ if the face $f$ goes through the edge $\ell$ in the
same direction, $-1$ if the face $f$ goes through the edge $\ell$ in the opposite
direction,  0 otherwise. Let us that notice that the total amplitude Eq. \eqn{gftampl}
is factorized as a product of face amplitudes and it reproduces the result that we
obtained for BF amplitudes in Eq. \eqn{bfampl}. We therefore have a positive answer to
the question we posed at the end of section 1, at least for simple models: we have a
field theory which generates the transition amplitudes for space-time geometries in the
absence of constraints. It can be easily shown that, when using the Peter-Weyl
decomposition for group variables we obtain an expression for the amplitude
\eqn{gftampl} in terms of 6j-symbols which is exactly the Ponzano-Regge model
\cite{ponzano-regge}.

In four dimensions  we have to implement the constraints described in section 1, at the
level of the discrete theory. There are various  proposals, the first being the
Barrett-Crane model \cite{BC} while the inclusion of the Barbero-Immirzi parameter led
to the EPRL/FK model \cite {EPRL,  FK}. Here  the Barrett-Crane model is recovered in
the $\gamma\rightarrow \infty$ limit.  Interestingly, recently new models have appeared,
with and without the Barbero-Immirzi parameter \cite{BaratinOriti1},
\cite{BaratinOriti2} which are based on the noncommutative algebra of flux variables.
Here we just sketch the EPRL/FK model.

\subsection{{Models of 4D gravity}}
The EPRL/FK model \cite{EPRL,FK} implements in two steps the constraints in Eq.
\eqn{constr} with a non trivial value of the Barbero-Immirzi parameter $\gamma$. In
order to describe the model we introduce $SU(2)$ coherent states \cite{Perelomov}
$$
|j,g \ket \equiv g |j,j\ket = \sum_m  |j,m \ket [R^{(j)}]^m_{j}(g) .
$$
with $|j,m>$ the eigenstates of the Lie algebra generators and $[R^{(j)}]^m_{j}(g)$ the
spin-j representation of the group element $g$. We have for the partition of unity
$$
1_j = \rd_j \int_{{\rm SU}(2)} dg \, |j,g \ket \bra j,g|= \rd_j \int_{G/H = S_2} dn \,
|j,n \ket \bra j,n|
$$
with $ |j,n \ket=g_n|j,j\ket $.
In  four dimensions  we use the  $SU(2)\times SU(2)$
coherent states \;\; $|j_+,n_+ \ket \otimes |j_-,n_- \ket$
$$
1_{j_+}\otimes 1_{j_-}  =  \rd_{j_+} \rd_{j_-}\int d n_+d n_- |j_+,n_+\ket\otimes
|j_-,n_-\ket \bra j_+,n_+|\otimes \bra j_-,n_-|
$$
In this language the constraints are implemented as \cite{FK}
\\
  $j_+/j_- = (1+\gamma) /(1- \gamma) $ \;and
\;\;$n_+=n_- = n$
$$
\gamma>1 \qquad \qquad j_{\pm}= \frac{\gamma \pm 1}{2} j,
$$
$$
\gamma<1 \qquad \qquad j_{\pm}= \frac{1\pm \gamma}{2} j.
$$
We consider now the propagator of the 4D BF theory, with gauge group $SO(4)$, which is a
natural generalization of the 3D propagator Eq. \eqn{prop}, when represented in the
coherent states basis
\beqa
\jj (g ; g') &=&\int_{SU(2)\times SU(2)} d u d v \prod_{f=1}^4 \sum_{j_{f+},j_{f-}}
\rd_{j_{f+}} \rd_{j_{f-}}\nonumber
\\&&\tr_{V_{j_{f+}}\otimes V_{j_{f-}}} \left( u g_f (g'_f)^{-1}
v^{-1}  {1_{ j_{f+}}\otimes 1_{ j_{f-}}}\right).
\eeqa
In each strand the identity  {$1_{j_+}\otimes 1_{j_-}$} is replaced by a  {projector
$T_j^{\gamma}$}
\be
T^{\gamma}_{ j}=   d_{j_+ + j_-} \bigl[  \delta_{j_{-}/j_{+} =
(1-\gamma) /(1+ \gamma) }  \bigr] \int d n  |j_+,n \ket\otimes
|j_-,n \ket \bra j_+,n | \otimes \bra j_-, n | .
\ee
which verifies
$
(T^{\gamma}_{ j})^2=T^{\gamma}_{ j}.
$
 Grouping the four strands of a line defines a $\TT$ operator
that acts separately and independently on each strand of the propagator:
\be
 \TT =
\oplus_{j_{f}}    \otimes_{f=1}^4
 T_{j_f}^{\gamma}
 \ee
so that the EPRL/FK propagator is
\be C= \jj \TT \jj.\ee
{The operator $C$  is symmetric. This implies that Feynman amplitudes are independent of
the orientations of faces and propagators.
 Since $\TT$ and $\jj$ do not commute, the propagator $C$ can
have non-trivial {\it spectrum} (with eigenvalues between 0 and 1). Moreover,  since
$\TT$ is a projector, the propagator $C$ of the EPRL/FK theory is bounded in norm by the
propagator of the $BF$ theory, as well as Feynman amplitudes.

To obtain the amplitude of a given graph ${\mathcal G}$ we combine the propagator and
the vertex expressions as in usual quantum field theory and  integrate
 over all $g,g'$ group variables (see \cite{KMRTV2010} for details). The total amplitude
 may be seen to be factorized as (the integral of ) a product of  face amplitudes
\begin{equation}
A_{\mathcal G} = \int \prod_{\ell \in L_{{\mathcal G}} } d u_\ell d v_{\ell} \prod_{ f
\in \mathcal{F}_{{\mathcal G}}}  {\mathcal A}_f
\end{equation}
with $\ell \in L_{{\mathcal G}}$ the edges of our graph, and $\mathcal{A}_f$ given by
\be
\mathcal{A}_f =\sum_{j_f\le \Lambda} d_{j_{f+}}d_{j_{f-}} \tr_{j_{f+}\otimes j_{f-}}
\prod_{\stackrel{a=1}{}}^p \left(h_{\ell_a,v_a} ^{\eta_{\ell_a f}}
 h_{\ell_{a}, v_{a+1}}^{\eta_{\ell_{a}f}}
T_{j_f}^\gamma\right)
 \ee
It can be seen that we recover the $SU(2)$ BF model  in the limit $\gamma\rightarrow 1$.
At this point we have all the ingredients of a quantum field theory. Specific graphs
have been computed (see for example \cite{KMRTV2010}) and their degree of divergence
analyzed. There is however no understanding on the perturbative expansion of the
partition function and a full renormalization group analysis is still lacking. A
modification of the model, which introduces colors for the fields has recently been
introduced. It allows for a better control  of the kind of topologies which are dually
associated to the graphs (see \cite {guraugft} and references therein).

\section{Noncommutative structures}
In this section we will only consider the three dimensional case, although some of the
results we will describe have been extended to the full 4d case \cite{BaratinOriti1,
BaratinOriti2}.

As we have seen, in three dimensions gravity is described by a BF theory with $SU(2)$
group and the group field theory  associated to its discretization is represented by the
Boulatov model, with action in Eq. \eqn{GFTaction} (with D=3).

We can define on the group manifold coordinate functions
\be
p^i=-i \tr g\sigma^i, \; \; i=1,..,3
\ee
where $\sigma$ are the Pauli matrices, and we parametrize $g\in SU(2)$ as $g= p^0 I +i
\sigma_i p^i$, with $(p^0)^2+ \sum_i (p^i)^2=1$. We indicate with $x_i$  the conjugate
variables  which live on the fibers of the cotangent bundle $T^*SU(2)$. The confusing
notation for the base and fiber coordinates is linked to the physical interpretation
from the gravity point of view. The canonical Poisson brackets on the cotangent bundle
are
\beqa
\{p^i,p^j\}&=&0\label{pp}\\
\{x_i,x_j\}&=&\epsilon_{ij}^k x_k \label{xx}\\
\{p^i,x_j\}&=&2(-\delta_{ij}\sqrt{1-|\vec p|^2}+\epsilon_{ijk}p^k) \label{ij}
\eeqa
They describe the dynamics of many interesting physical systems, as  for example the
dynamics of the rigid rotor with $x_i$ associated to the angular momentum components and
$p^i$ to the orientation of the rotor, or, when generalized to  field theory, the
Poisson algebra of currents for the principal chiral model.

As a group $T^*SU(2)$ is the semidirect product of $SU(2)$ and the abelian group
$\mathbb{R}^3$, with Lie algebra the semidirect sum represented by
\beqa
\left[J_i,J_j\right]  &=& \epsilon_{ij}^k J_k \label{JJ}\\
\left[P_i,P_j\right] &=& 0 \label{PP}\\
\left[J_i,P_j\right] &=&\epsilon_{ij}^k P_k. \label{JP}
\eeqa
The non-trivial Poisson bracket on the fibres of the bundle, \eqn{xx},  is usually
understood in terms of coadjoint action of the group $SU(2)$ on its dual algebra
$\mathcal{L}^*=(\mathbb{R}^3)^*\simeq \mathbb{R}^3$ and it reflects the non-triviality
of the Lie bracket \eqn{JJ}\footnote{The Lie algebra generators $J_i$ are identified
with the linear functions on the dual algebra}.

The question arises whether the non-trivial Poisson bracket on
$\mathcal{F}(\mathcal{L}^*)$ may be quantized yielding  a noncommutative star-product in
the spirit of deformation quantization. This is relevant to our problem because, in the
BF picture,  the group variables are associated to the holonomies while the   $x_i$
variables are associated to the triad components \cite{bo2010}.

 I am aware of essentially
two different answers and it is not clear at the moment what is the relation among them.

The first approach consists in regarding the algebra $\mathcal{F}(\mathcal{L}^*)$ as a
subalgebra of the algebra of quadratic functions on $\mathbb{R}^4$. This is known as the
classical Jordan-Schwinger map or symplectic realization. For details we refer to the
existing literature \cite{MMVZ94, GLMV01}. Once such an immersion is realized, one can
use the Moyal product on $\mathcal{F}(R^4)$ or variations of it (see for example
\cite{HLS} where the Voros product has been used and \cite{Li2011} for a recent
application) to induce a star product on $\mathcal{F}(\mathcal{L}^*)$. It can be shown
that the subalgebra is closed under the product. The symplectic realization of the
coordinate functions $x_i$ is
\be
x_i=\bar z^a\sigma^{ab}_i z^b, \;\;\; x_0=\bar z^a\delta^{ab} z^b
\ee
with $x_0=|\vec x|$ in the kernel of the projection, $a,b\in{1,2}$ and we have made the
identification $\mathbb{R}^4\simeq \mathbb{C}^2$ with canonical symplectic structure
\be
\{\bar z^a,z^b\}=i \;.
\ee
Let us notice that similar realizations of a 3d Lie algebra as Poisson subalgebra of
quadratic functions on $\mathbb{R}^4$ have been derived for all the 3d Lie algebras
\cite{MMVZ94}.  The Moyal star product
\be
\phi\star_M\psi (\bar z, z)= \phi \exp\left(\frac{\theta}{2} \overleftarrow{\del}_{z_a}
\overrightarrow{\del}_{\bar z_a}-\overleftarrow{\del}_{\bar z_a}
\overrightarrow{\del}_{z_a}\right)\psi
\ee
induces in $\mathcal{F}(\mathcal{L}^*)$ the product
\be
(x_i\star_M \phi) (x) = \left\{x_i
-i\frac{\theta}{2}\epsilon_{ijk}x_j\del_k-\frac{\theta^2}{8}[(1+x\cdot
\del)\del_i-\frac{1}{2}x_i\del\cdot\del]\right\}\phi(x)
\ee
which implies for coordinate functions
\be
x_i\star_M  x_j= x_i \cdot x_j + i\frac{\theta}{2}\epsilon_{ijk}x_k
-\frac{\theta^2}{8}\delta_{ij}
\ee
Once again, similar expressions exist not only for $SU(2)$ but for all 3d cases
\cite{GLMV01}. If we replace the Moyal product with the  Voros product
\be
\phi\star_V\psi (x)= \phi \exp\left(\theta \overleftarrow{\del}_{z_a}
\overrightarrow{\del}_{\bar z_a}\right)\psi
\ee
we get instead
\be
x_i\star_V  x_j= x_i \cdot x_j + {\theta}(|\vec x|+ i \epsilon_{ijk}x_k)
\label{starvoros}
\ee
with $x_0=|\vec x|$. Let us point out that indeed a whole family of star products can be
derived, corresponding to different ordering choices in the quantization procedure on
the plane. These products,  Moyal and Voros products being just two representatives, are
characterized by being translation invariant, therefore  reproducing  the same star
commutator \cite{GLV}.

 The second approach consists in defining the star product for
$\mathcal{F}(\mathcal{L}^*)$ in terms of a group Fourier transform
\be
\tilde\phi(x)=\int dg \phi(g) e^{\tr (g\vec\sigma) \cdot \vec x}
\ee
with
\be
e^{\tr (g_1\vec\sigma) \cdot \vec x}\star_F e^{\tr (g_2\vec\sigma)\cdot \vec
x}:=e^{i\tr (g_1 g_2\vec\sigma)\cdot \vec x}
\ee
 It was first introduced in \cite{LivineFreidel}, then further investigated in \cite{majid,noui}.
 It has been adapted to GFT in \cite {bo2010} and recently extended to the four
dimensional case in \cite{BaratinOriti1,BaratinOriti2}. We refer to the literature for a
proper definition of the product, limiting ourselves to observe that the induced star
product among coordinates does not coincide with the Moyal-induced one. We have instead
\be
x_i\star_F x_j= x_i \cdot x_j + i\kappa\epsilon_{ijk}x_k \label{Fstar}
\ee
with $\kappa$ a suitable constant, needed to fix the dimensionality. The interesting
feature of this product is that it naturally arises in the GFT action for the Boulatov
model, when we pass to the Fourier transform \cite{bo2010}.

It would be interesting to understand what is the relation between all these products,
given that they realize the same commutation relations
\be x_i\star x_j-x_j\star x_i= i \epsilon_{ijk} x_k
\ee
up to multiplicative constants. In particular we would like to  understand whether the
Fourier-related  star product in Eq. \eqn{Fstar} may be  induced from one of the
translation invariant
 star products on the algebra $\mathcal{F}(\mathbb{R}^4)$, via symplectic
realization.

To conclude this section on noncommutative structures in GFT let me speculate on the
issue of recovering the cosmological term in the GFT action.

It is known that, at the level of spin-foam amplitudes, the cosmological constant is
taken into account on replacing the group $SU(2)$ with its quantum analogue  $SU_q(2)$.
This is the Turev-Viro model \cite{Turaev}. On the other hand, at the classical level,
it is known since the famous paper of Witten \cite{witten} that the cosmological
constant is easily introduced in the 3D action of gravity if one regards gravity with
zero cosmological constant as a Chern-Simons theory for the Poincar\'e gauge group
$ISO(2,1)$.  Then one deforms the algebra of $ISO(2,1)$ into a fully nonabelian one
($SO(3,1)$ or $SO(2,2)$, depending on the sign of the cosmological constant).

If we look back at the starting Poisson algebra of coordinate functions Eqs.
\eqn{pp}-\eqn{ij} and its Lie algebra counterparts Eqs \eqn{JJ}-\eqn{JP}, we realize
that this amounts to modify Eq. \eqn{PP} in the Lie algebra and dually Eq. \eqn{pp} in
the Poisson algebra. This makes $SU(2)$ into a Lie-Poisson group. Its full quantization
should give the desired quantum group and allow to recover the cosmological term at the
GFT level. We shall come back to this issue in a separate publication.

\end{document}